\documentclass[cameraready]{Interspeech}
\title{\textit{AudioNoisePrints}: Model-free audio watermarking using spatial correlation in flow matching TTS}

\author[affiliation={1,2}, orcid=0009-0006-4437-7418]{Timothy Tin-Long}{Tse}
\author[affiliation={2}, orcid=0000-0002-7849-1060]{Jian}{Zhu}
\author[affiliation={1}, orcid=0000-0001-6552-5324]{Aidan}{Pine}
\author[affiliation={1}, orcid=0000-0002-7886-439X]{Mengzhe}{Geng}

\address{
    $^1$ National Research Council Canada, Canada
    $^2$ University of British Columbia, Canada
}

\email{timothytse1997@gmail.com, jian.zhu@ubc.ca, aidan.pine@nrc-cnrc.gc.ca, mengzhe.geng@nrc-cnrc.gc.ca}

\keywords{Audio Watermarking, Speech Synthesis}

\usepackage{comment}
\usepackage{mathtools}

\newcommand\sreff[1]{\S\ref{#1}}

\begin{document}

\maketitle

\begin{abstract}
We present \textit{AudioNoisePrints}, a \textit{training-free} watermarking pipeline for flow matching and diffusion TTS models, which requires minimal extra computation during inference and does not require retraining the TTS model or reducing the generation quality. We exploited the fact that there are strong correlations between the initial Gaussian noises and the generated outputs in diffusion and flow matching models, such that a simple cosine correlation between the initial noise and the generated output can be used to perform watermaking. Moreover, we train a lightweight detector on top for more aggressive augmentations. 
Our method outperforms \textit{AudioSeal}, a strong baseline for audio watermarking under strong augmentations. We experimented on  \textit{F5TTS} and other TTS and vocoder models, and concluded that they all exhibit similar \textit{spatial correlation} properties, suggesting our watermarking scheme can be used for more flow-matching TTS models and even vocoders in the future.
\end{abstract}

\section{Introduction}

Flow matching models \cite{lipman2023flow} and diffusion models \cite{NEURIPS2020_4c5bcfec} have anable impressive progress in speech synthesis, from multi-speaker Text-to-Speech models like MatchaTTS \cite{mehta2024matchattsfastttsarchitecture}, E2TTS \cite{eskimez2024e2ttsembarrassinglyeasy}, and F5-TTS \cite{chen2025f5ttsfairytalerfakesfluent}, to vocoders like DiffWave \cite{kong2021diffwaveversatilediffusionmodel} and WaveFM \cite{luo2025wavefmhighfidelityefficientvocoder}. While they are praised for their capability in high-fidelity speech synthesis, very little is known of their properties beyond generation quality.   

In recent years, researchers have tried to analyze the relationship between the intermediate latent representations of diffusion and flow matching (FM) models and the semantics of the generated data. Recent results showed the correlation might be straightforward: the latent encodings (originating noise) somewhat resemble the generated image patterns \cite{staniszewski2025againrelationnoiseimage, goren2025noiseprintsdistortionfreewatermarksauthorship}. More specifically, the generated image has a much stronger correlation with the originating noise than any other randomly generated noise (see \sreff{background} for elaboration). This property allows a simple method of watermarking (named \textit{NoisePrints} \cite{goren2025noiseprintsdistortionfreewatermarksauthorship}): deterministically creating a set of initial noise for generation and using the correlation between the known noise and generated output as the metric to determine if the image originated from the noise.
This is especially important for watermarking, as it not only prevents the watermarking process from reducing the quality of the generated image \cite{choi2025visual}, but it also does not require retraining the flow-matching model or an external detector. In this way, the approach is `model-free'.

In the original work for \textit{NoisePrints}, only one specific type of model was explored: {a latent diffusion model which generates images via a VAE. To our knowledge, we are the first to investigate this approach on diffusion or flow matching (FM) based speech-synthesis models. In this work, our main contributions can be summarized as follows: 

\begin{figure*}[t]
    \centering
    \includegraphics[width=0.70\textwidth]{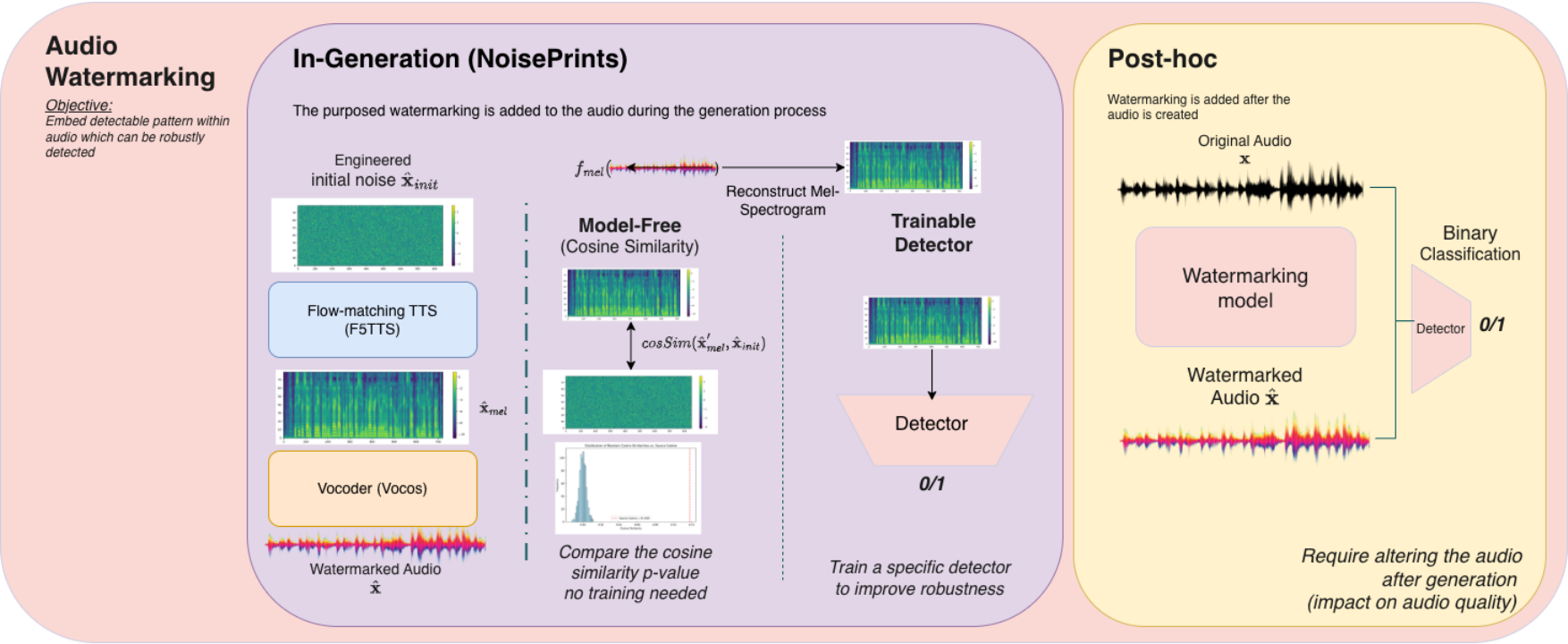}
    \caption{General overview of our approach (\textbf{\textit{AudioNoisePrints}}) comparing to the commonly used post-hoc watermarking schemes. As shown above, the post-hoc watermarking model required altering the audio after it is generated, which creates computation overhead and alters the quality of the audio. Our method simply changes the initial noise $\hat{\textbf{x}}_{init}$ during generation (without altering the quality of the generated audio), and uses either the model-free option (calculating the cosine similarity of the audio-Mel-spectrogram and the initial noise) or the more robust detector option to determine if the audio is in fact watermarked.}
    \label{fig:audiowm_overview}
\end{figure*}
\begin{itemize}

    \item Our experiments with F5-TTS and MatchaTTS shows our new watermarking scheme achieved better robustness under strong augmentation when comparing with state-of-the-art post-hoc deep watermarking models; without the need of retraining the TTS model or introducing extra compute during inference time. Our method also circumvents the trade-off between ``\textit{quality} vs \textit{robustness}''
    \item We further extend the original approach by introducing an external detector, which again achieves better result against the strong baseline model while maintaining the advantages of \textit{NoisePrints}'s methodology (no overhead during generation and no impact on generation quality).
    \item We show that other flow matching (FM) and diffusion TTS and vocoder models also exhibit strong correlations between originating noises and generated data as described by \cite{staniszewski2025againrelationnoiseimage, goren2025noiseprintsdistortionfreewatermarksauthorship}, proving more TTS and vocoder models can be benefited with our simple watermarking approach.

\end{itemize}

\section{Background}\label{background}

\subsection{Diffusion and Flow-matching}\label{DandFM}
Diffusion models and flow matching (FM) models are a family of iterative generative models with a similar process of reversing Gaussian noise into the target data distribution \cite{ho2020denoisingdiffusionprobabilisticmodels, lai2025principlesdiffusionmodels}. Consider a data point $x_0$ (could either be an image, a Mel-spectrogram, or audio frame) and a Gaussian noise $x_L$ with intermediate steps $i\in[0, L]$. For the diffusion model $\epsilon_\phi(x_i, i)$, we have the following iterative equation (see Lemma 2.2.2 and Equation 2.2.9 in \cite{lai2025principlesdiffusionmodels}): 

$$x_{i-1} = \frac{1}{\alpha_i}\bigg(x_i - \frac{1-\alpha_i^2}{\sqrt{1-\bar{\alpha}_i^2}}\epsilon_\phi(x_i, i)\bigg)$$

$\alpha_i$ and $\beta_i$ are predetermined constants with $\alpha_i \coloneq \sqrt{1 - \beta_i^2}$ and $\beta_i \in (0, 1)$, and $\bar{\alpha}_i \coloneq \prod_{j=0}^{i}\alpha_i$. On the other hand, for flow matching models $v_\phi(x_i, i)$, we have 
$$x_{i-1} = x_i + v_\phi(x_i, i)dt_i$$

Where $dt_i$ are, again, predetermined constants. In both cases, we can see that for each step of $i$, the model output of diffusion ($\epsilon_\phi(x_i, i)$) and flow-matching ($v_\phi(x_i, i)$) models are trained to guide the Gaussian noise from $x_L$ all the way to the data distribution of $x_0$, acting similar to a vector field in the target data space (note that we only consider the case for using the Euler ODE solver, which is the only solver used in all experiments in this paper).

In recent developments, there have been multiple findings on the relationship between $x_L$ and the generated $x_0$ via diffusion model, namely, the spatial relations between the two \cite{staniszewski2025againrelationnoiseimage, goren2025noiseprintsdistortionfreewatermarksauthorship}. They found that the cosine similarity between the generated $x_0$ and originating noise $x_L$ is significantly higher than that of other randomly sampled noise $x'_0 \sim \mathcal{N}(0,I)$. This resulted in a higher Pearson correlation between each pixel in $x_0$ and $x_L$. It seems the patterns in the generated image (or latent) are also present in the original noise.

\subsection{Audio Watermarking}
Deep audio watermarking algorithms \cite{roman2024proactivedetectionvoicecloning, chen2024wavmarkwatermarkingaudiogeneration, roman2024latentwatermarkingaudiogenerative, liu2023detectingvoicecloningattacks, liu2024grootgeneratingrobustwatermark} strive to inject robustly detectable yet imperceptible information within audio. Most current approaches fall under the following two categories.

\subsubsection{Post-hoc watermarking}
 Applied on the already generated audio (hence the name ``post-hoc''), these models are plug-and-play in the sense that you can simply download an open-source watermarking model and implement it in your pipeline. However, recent research has shown that using open-source watermarking schemes is prone to \textit{overwriting attacks} \cite{yao2025mineoverwritingattacksneural}, where the attacker simply has to detect which watermarking model is used in the audio and proceed to overwrite it with their own watermark (using the same open-source model). Post-hoc watermarking also requires extra computational resources and latency in the audio generation pipeline. Moreover, post-hoc watermarking requires the audio to be altered; the more audibly altered the audio is, the more robust the watermark becomes to malicious attacks (harder to erase), but this also means there exists a fundamental trade-off between audio quality and robustness.

\subsubsection{In-model watermarking}
On the other hand, in-model watermarking \cite{zhao2025traceablettswatermarkfreetts, roman2024latentwatermarkingaudiogenerative} tries to introduce the watermark during the generation process, which usually requires retraining the generative model (for example, \cite{zhao2025traceablettswatermarkfreetts} retrain the entire TTS model in order to make it detectable). While these methods avoid some of the shortcomings of post-hoc watermarking, such as extra computation and vulnerability to overwriting attacks, they require many more computational resources to train, and the generation quality might also be affected after the in-model watermarking training is conducted, as one cannot simply ``add" in-model watermarking on top of their existing audio generation pipeline.

Notably, \textit{NoisePrints} \cite{goren2025noiseprintsdistortionfreewatermarksauthorship} is an in-model watermarking algorithm for latent image generation models. 
The \textit{NoisePrints} watermarking method uses the property of spatial similarity between noise and output mentioned in section \ref{DandFM} by creating a set of predetermined initial noise $x_0^{special}$ (from a known hash). By directly comparing that initial noise and the image's latent $x'_L$ (encoded by the same VAE encoder) via cosine similarity $d_{sim} = cosSim(x_0^{special}, x'_L)$, they can determine if that image was in fact generated from that particular noise via comparing the similarity score with an empirical threshold $\tau$. In short, if $d_{sim} > \tau$, the image will be determined as having been generated from that particular noise.

In this work, we are able to leverage such properties of diffusion and flow matching models to create an in-model pipeline which \textit{does not require retraining the audio generation model}, while also preserving most (if not all) of the generation quality.

\section{Method}\label{sect:method}


In this section we will discuss the ways in which we adapt the \textit{NoisePrints} method from the image domain to audio.

\begin{figure*}[tbp]
    \includegraphics[width=0.24\linewidth]{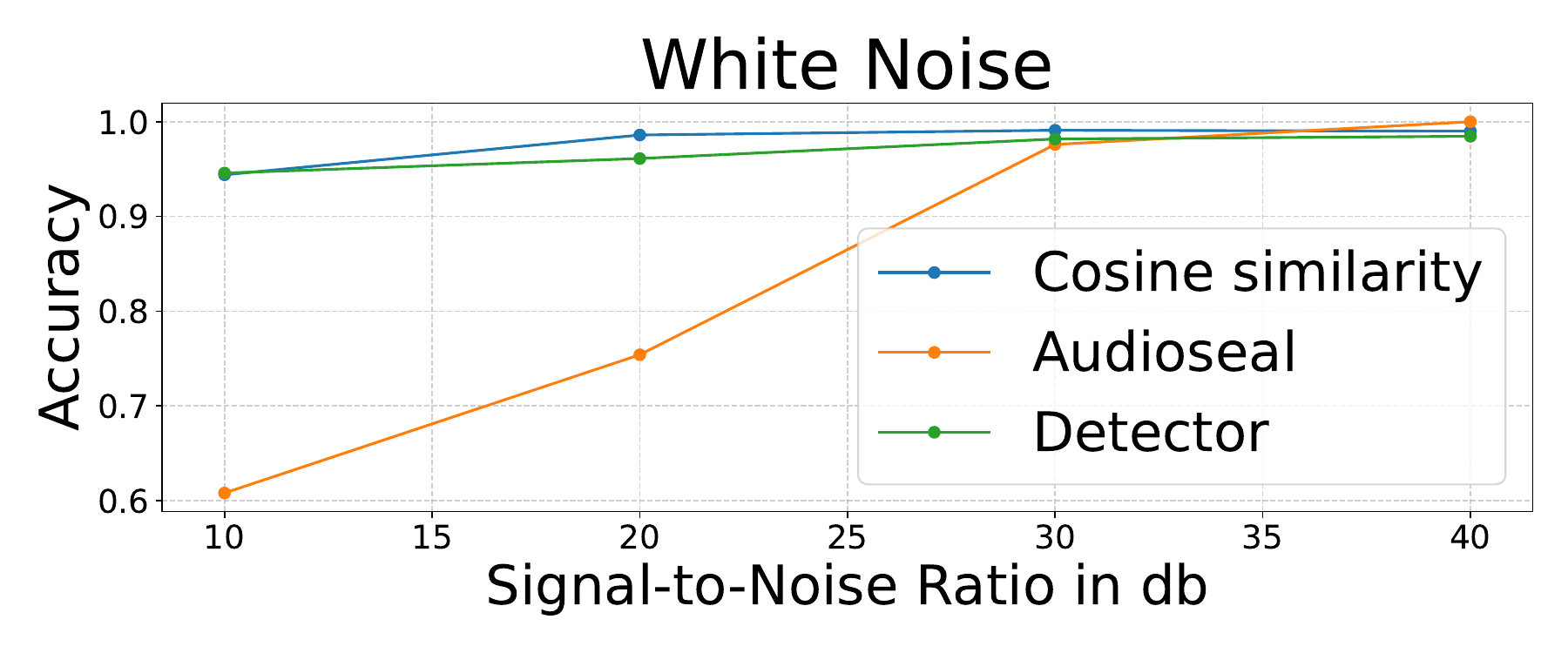}
    \includegraphics[width=0.24\linewidth]{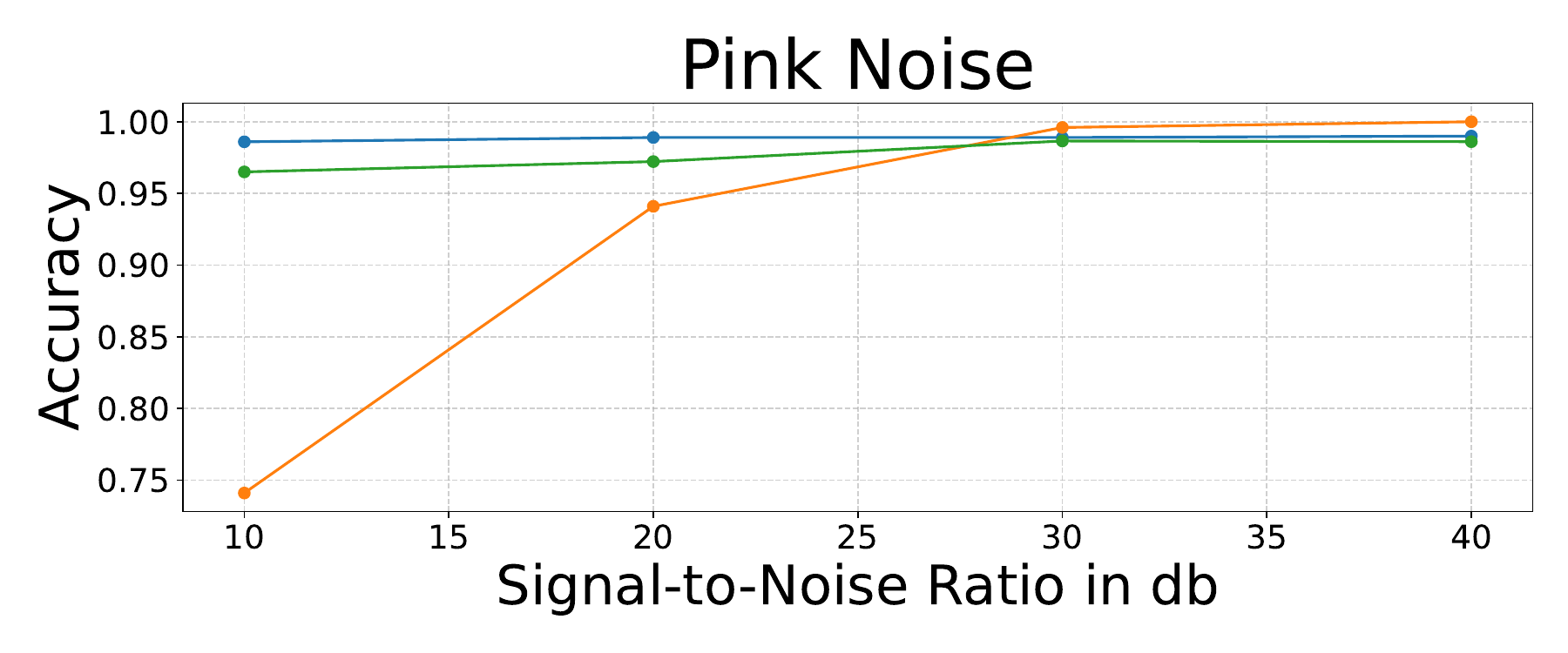}
    \includegraphics[width=0.24\linewidth]{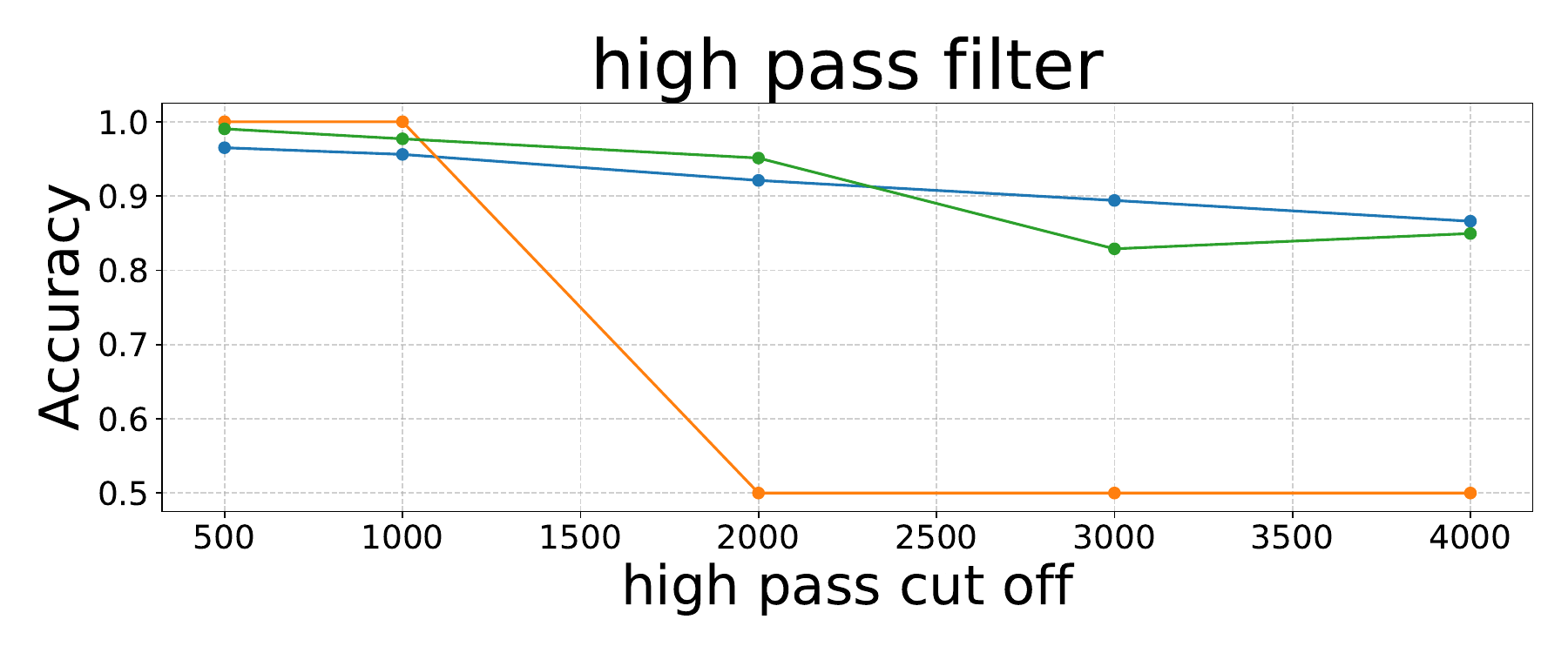}
    \includegraphics[width=0.24\linewidth]{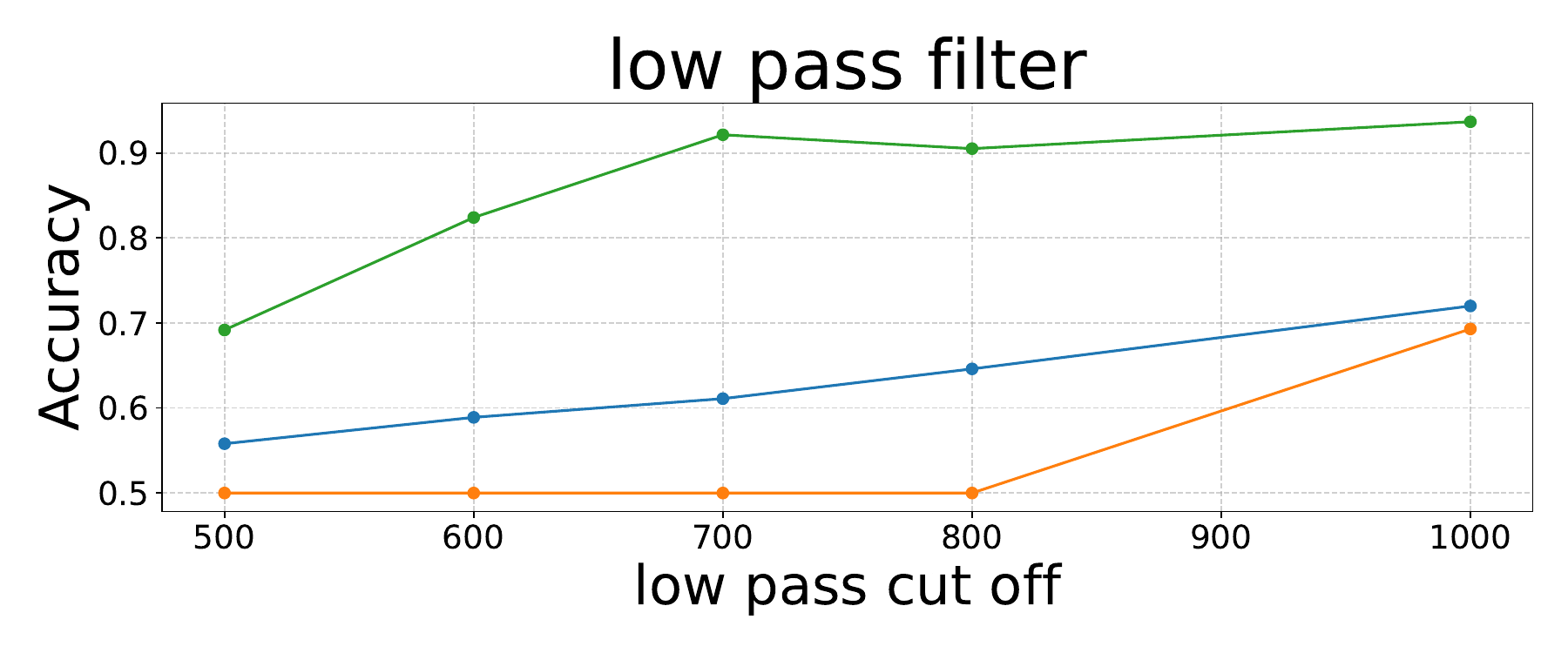}
    \includegraphics[width=0.24\linewidth]{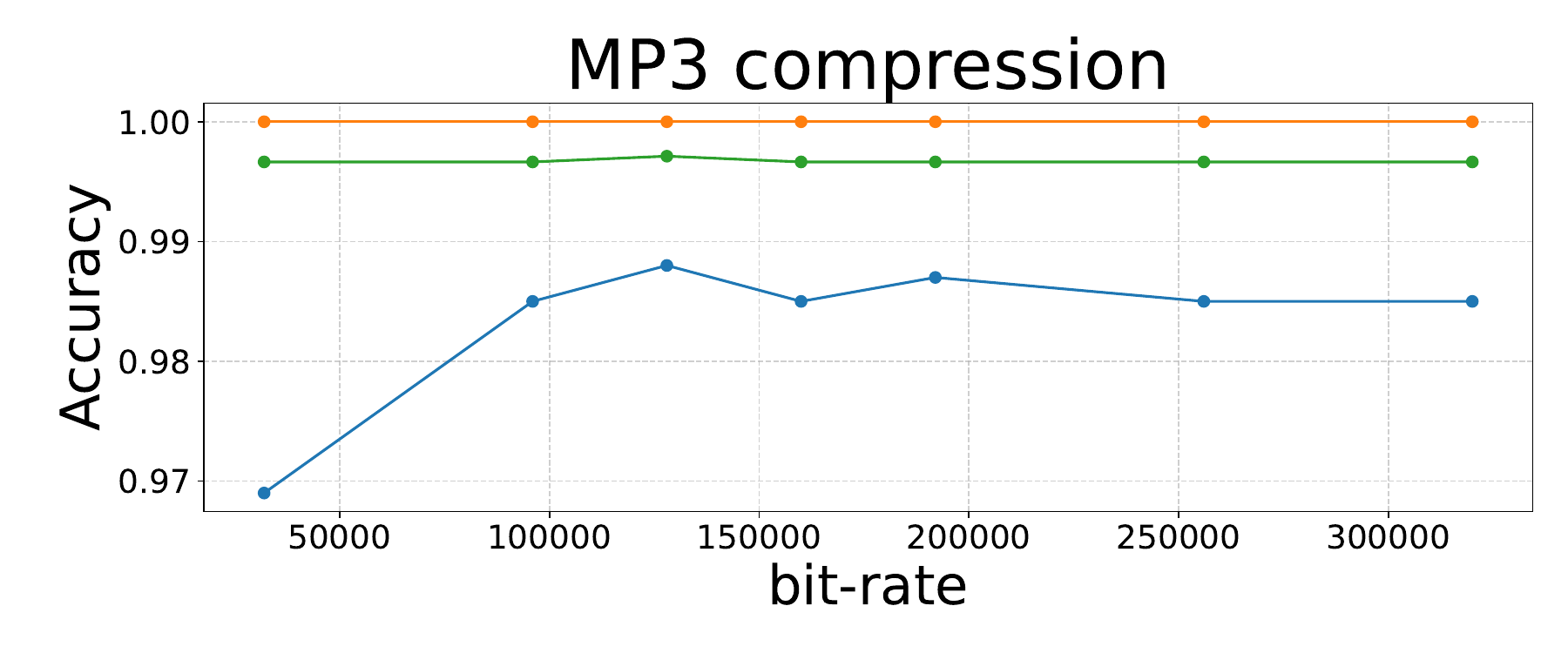}
    \includegraphics[width=0.24\linewidth]{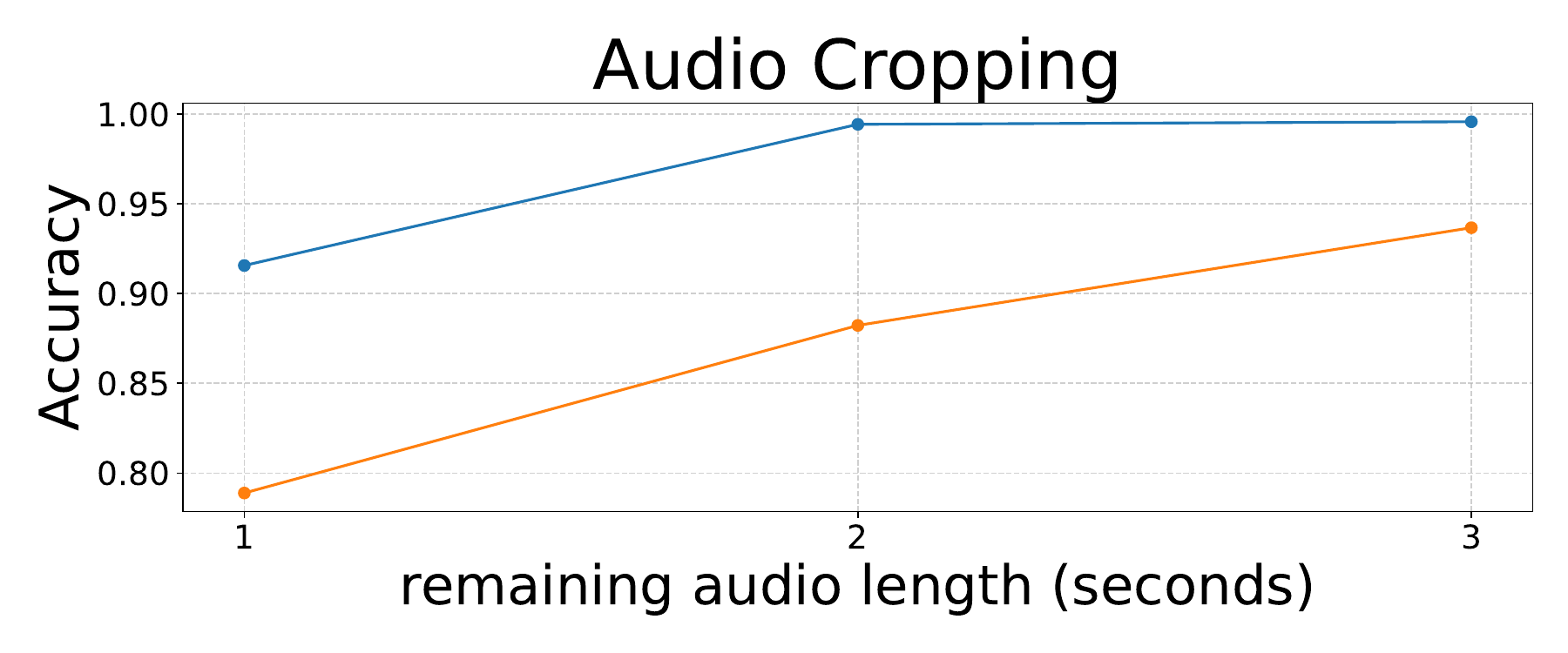}
    \includegraphics[width=0.24\linewidth]{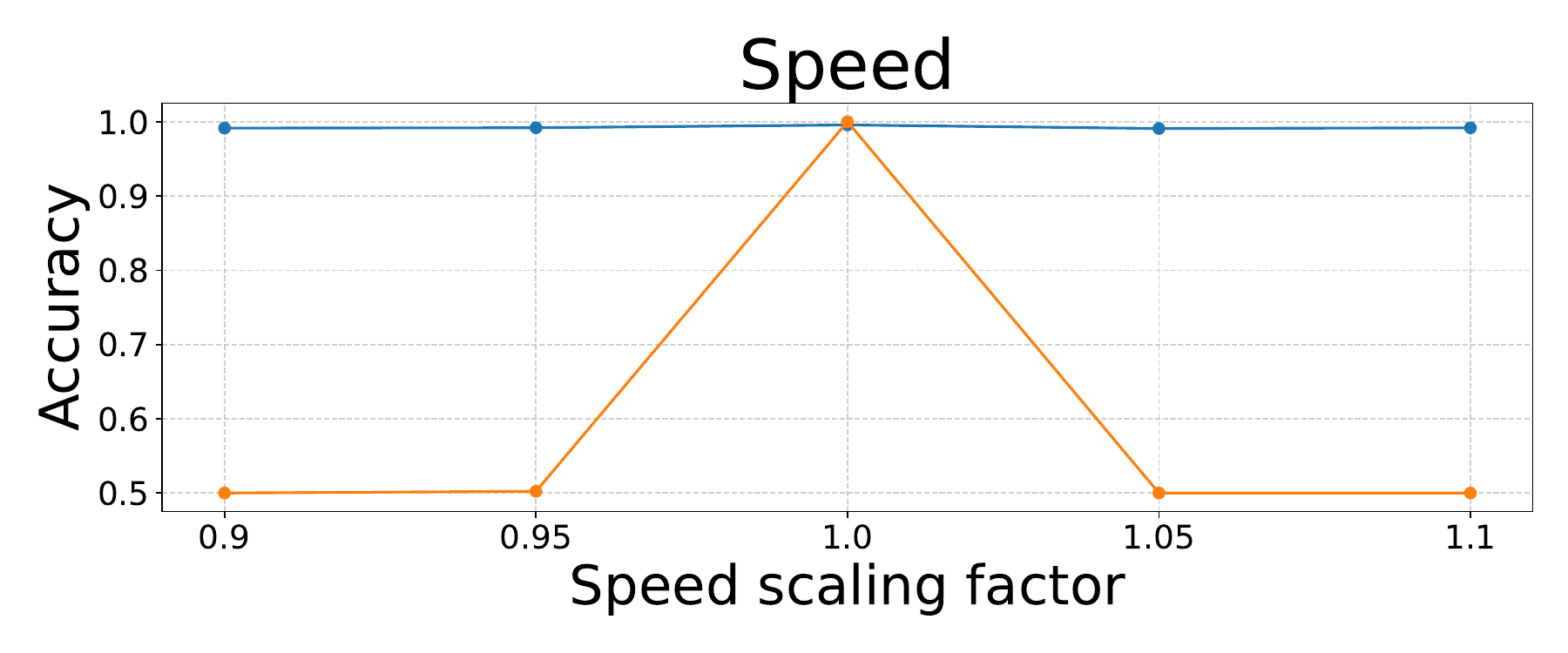}
    \includegraphics[width=0.28\linewidth]{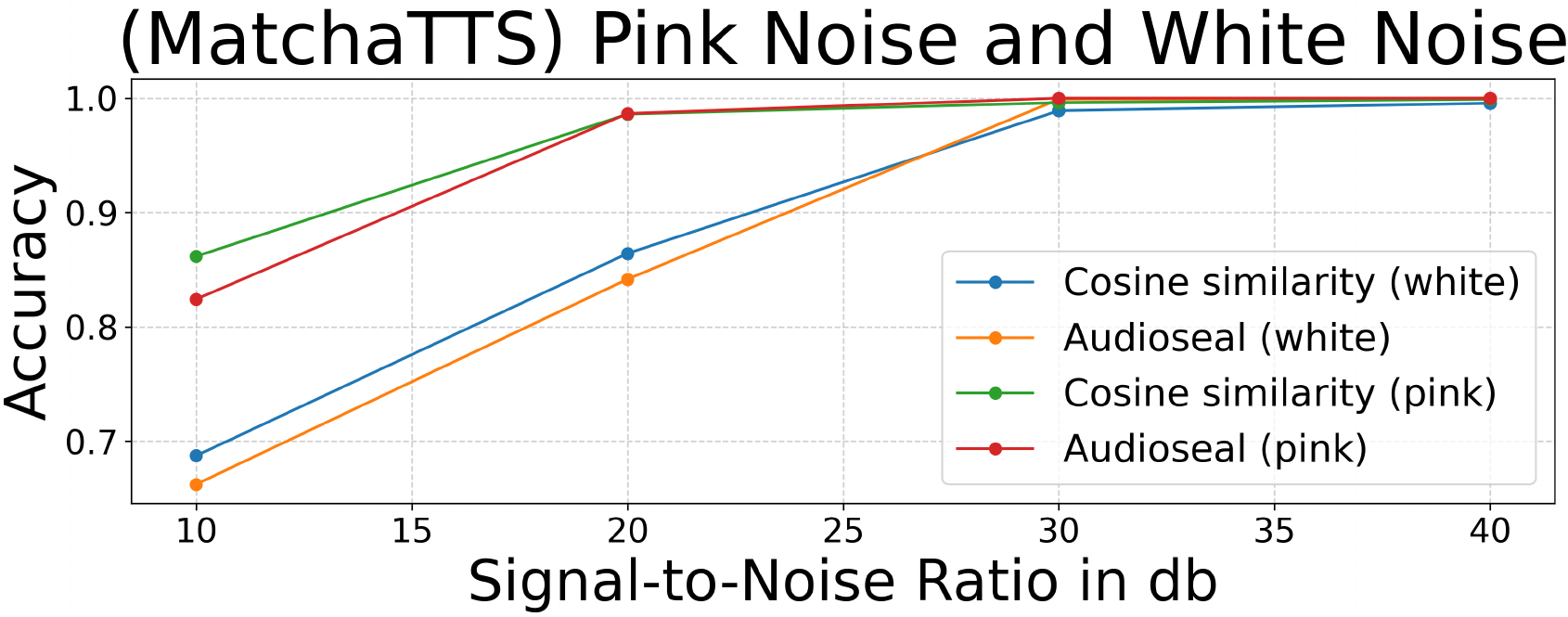}
    \includegraphics[width=0.28\linewidth]{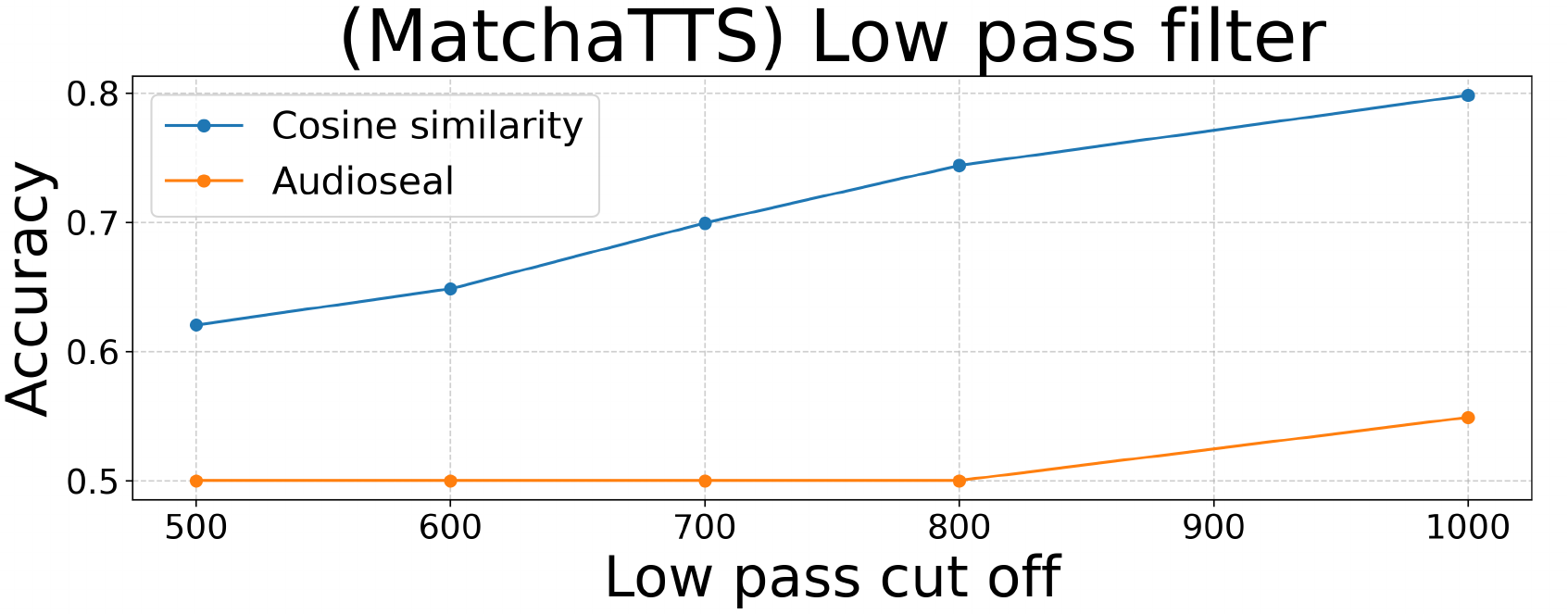}
    \medskip
    \centering
    \includegraphics[width=0.28\linewidth]{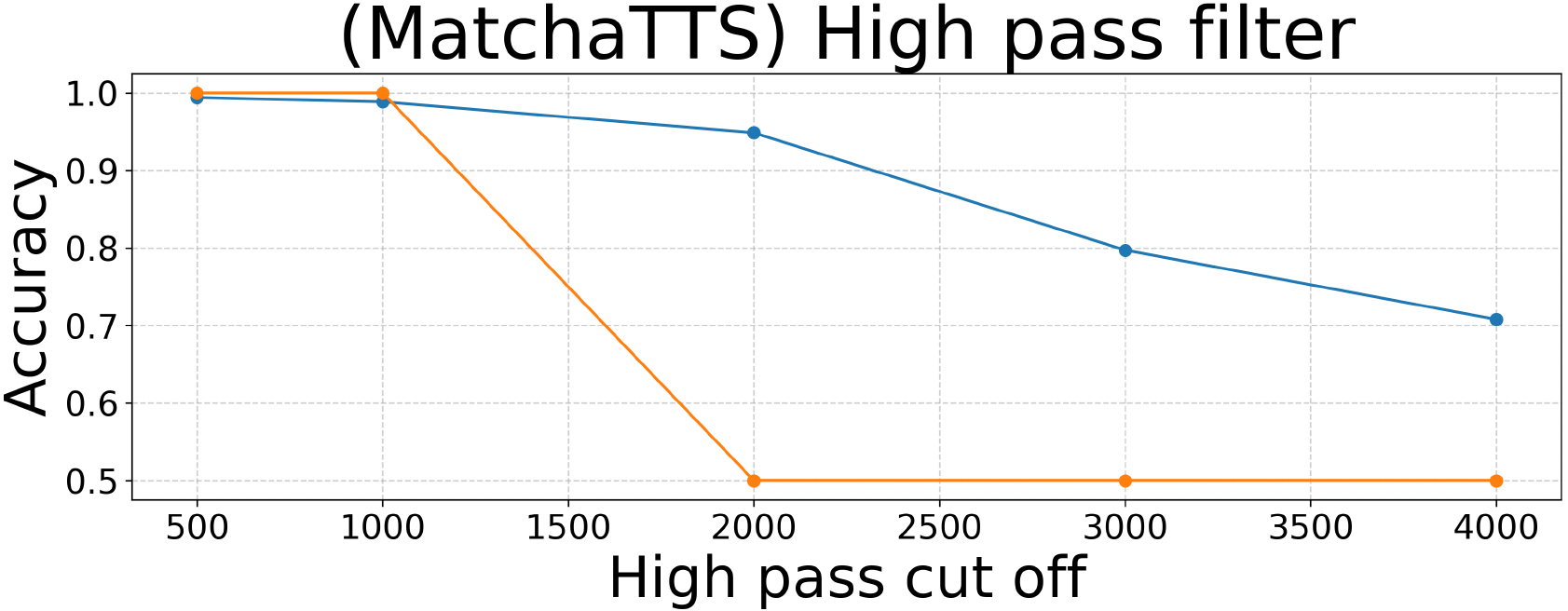}
    \caption {Accuracy comparing the AudioSeal and our approaches. While AudioSeal has a higher top-end score (1.0 accuracy), under strong augmentation, our method (using either the model-free \textit{Cosine similarity} approach or the \textit{Detector} approach) significantly outperforms the post-hoc watermark baseline. Especially in the speed augmentation, where AudioSeal fails even witha  1 percent change in stretching.}
    \label{fig:F5-TTS_audioseal_comparison}
\end{figure*}

\subsection{Adoption of NoisePrints in audio domain}\label{sect:NoisePrintsMethod}

We define the distance between the original noise ($x_0$) and the generated audio and/or Mel-spectrograms $x_L$ as $d_{org} = f_{dist}(x_0, x_L)$, with $f_{dist}$ being distance functions such as cosine-similarity. Unlike the original approach, the distance $d_{org}$ is not calculated in the latent space but rather in the audio or Mel-spectrogram domain. Instead of using a fixed threshold $\tau$ like in \cite{goren2025noiseprintsdistortionfreewatermarksauthorship}, we calculate the empirical p-value by comparing the distance between Mel-spectrogram and original noise $d_{org} = f_{dist}(x_0, x_L)$ and the distances between the same audio but with $N$ (500 in all of our experiments) randomly generated noise, i.e $d_{rand} = f_{dist}(x_{rand}, x_L)$. A threshold $\tau_p$ is set for the minimum p-value required for a audio to be considered ``watermarked". 

\subsection{The \textit{Detector} approach} 
Adapting \textit{NoisePrints} from the image to audio domain also poses new challenges. In TTS, an attack in the frequency domain could significantly reduce our detection accuracy. This raises the question: How robust is this spatial correlation in the audio Mel-spectrogram space? Moreover, as audio does not have a fixed size like a predetermined latent, simple attacks such as cropping and speed augmentation can cause the watermarking recovery to become significantly more complex, requiring algorithms that use sliding windows and dynamic time warping.

While we can still use cosine similarity, we devise a new watermarking pipeline to reduce these complexities during detection. We name it the \textit{Detector} approach. Given a special noise ($\hat{x}_{init}$ in Figure \ref{fig:audiowm_overview}), we train an external convolutional neural net with the objective of binary classification (if that audio is generated from the noise $\hat{x}_{init}$ or not, via a specific TTS diffusion/FM model). In order to increase robustness to combat the issue mentioned above, we apply data augmentation methods (including compression codecs like MP3 and AAC). We synthesize data from the TTS models with two categories: 1) generated from $\hat{x}_{init}$, and 2) noise $x_0$ being randomly sampled. To make the noise prediction even more reliable to combat aggressive attacks such as cropping, we sample the noise $\hat{x}_{init}$ periodically with a fixed length of 100 frames in Mel-space, i.e. for every audio with length $L$, we repeat $\hat{x}_{init}$ until the final length exceeds $L$.

We want to emphasize that while the \textit{Detector} approach requires additional training, the biggest benefit of a model-free approach still persists, namely we do not need to retrain the TTS model (which could lead to quality degradation). The detector approach is only used here to increase inference speed and reduce complexity.

\begin{table}[tbp!]
    \small
    \centering
    \begin{tabular}{lccccc}
        \toprule
        \textbf{$\tau_p$} & \textbf{0.80} & \textbf{0.85} & \textbf{0.90} & \textbf{0.95} & \textbf{1.00} \\
        \midrule
        \textbf{CoSim} & \textbf{0.905} & \textbf{0.930} & 0.949 & \textbf{0.975} & \textbf{0.988} \\
        \textbf{L1}    & 0.392 & 0.422 & 0.450 & 0.476 & 0.497 \\
        \textbf{L2}    & 0.405 & 0.424 & 0.448 & 0.474 & 0.497\\
        \textbf{Dot}    & 0.885 & 0.922 & \textbf{0.955} & 0.972 & 0.986 \\
        \bottomrule
    \end{tabular}
    \caption{Performance comparison across similarity thresholds and distance functions (F5-TTS). }
    \label{tab:similarity_thresholds}
\end{table}

\begin{table}[tbp!]
\centering
\begin{tabular}{lccc}
\toprule
& \multicolumn{3}{c}{\textbf{F5-TTS}} \\
\cmidrule(lr){2-4}
\textbf{Augmentation} & \textbf{CosSim} & \textbf{Detecor} & \textbf{AudioSeal} \\
\midrule
AAC compression & 0.987 & 0.9947 & 1.0 \\
Echo & 0.504 & 0.499 & 1.0 \\
\bottomrule
\end{tabular}
\caption{All three methodologies have near 1.0 accuracy AAC (mp4) compression. Yet the current methodology did not perform as well with the ``echo" augmentation.}
\label{table:compression_result}

\end{table}

\subsection{Experiments}\label{sec:models}

\begin{figure*}[tbp]
    \centering

    \includegraphics[width=0.30\linewidth]{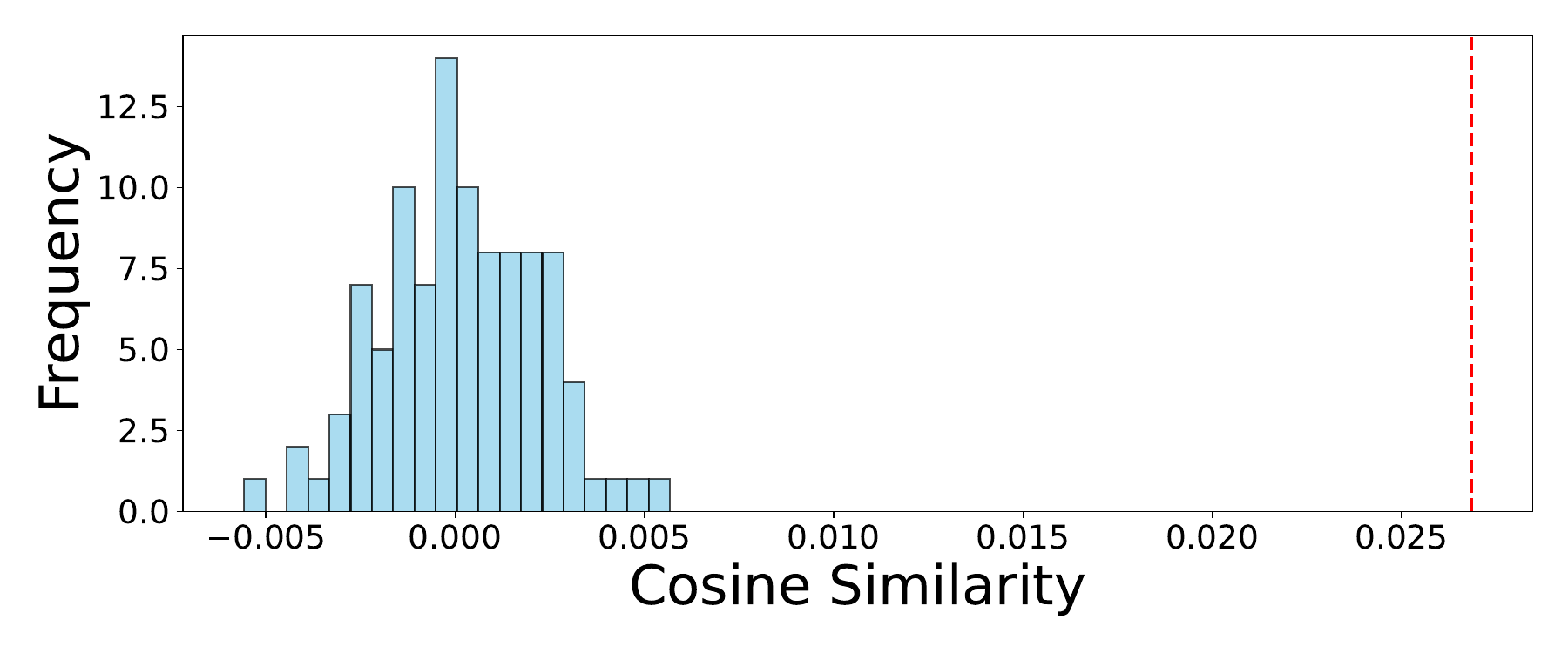}
    \includegraphics[width=0.30\linewidth]{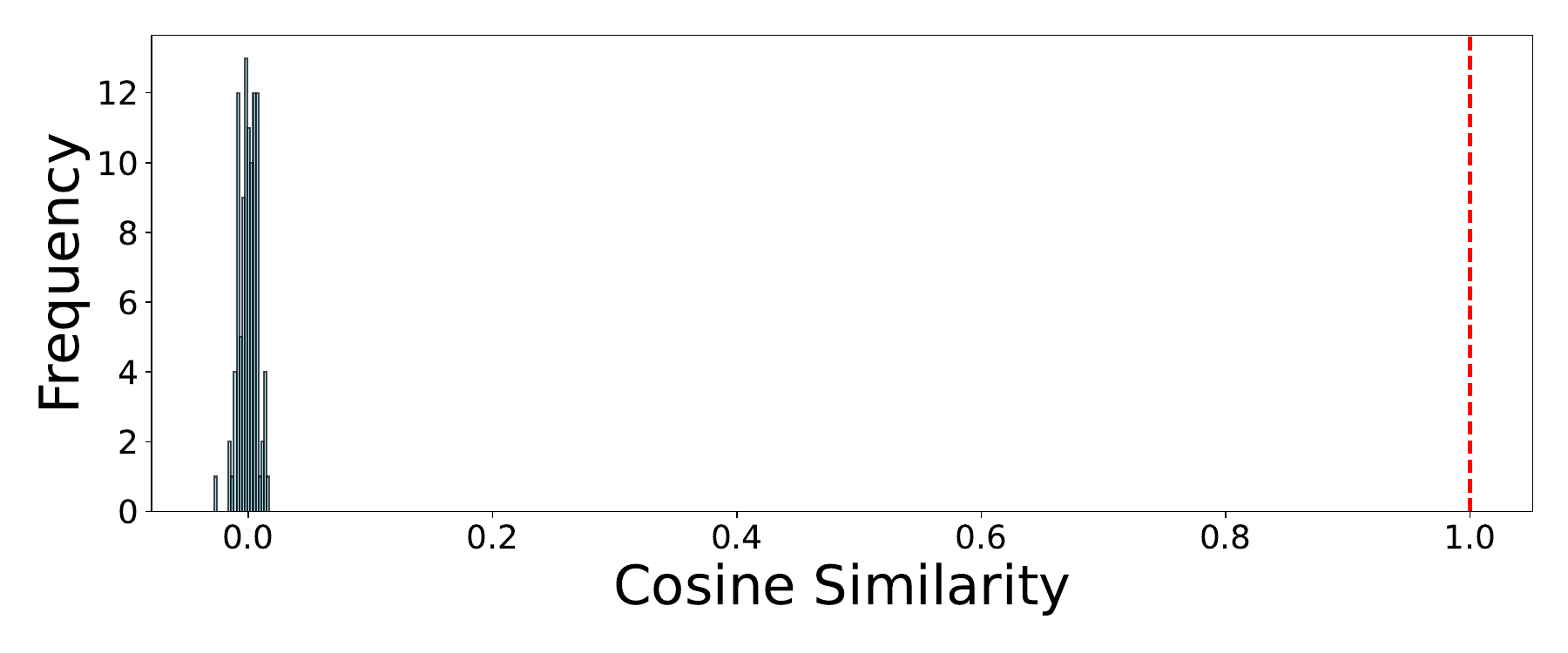}
    \includegraphics[width=0.30\linewidth]{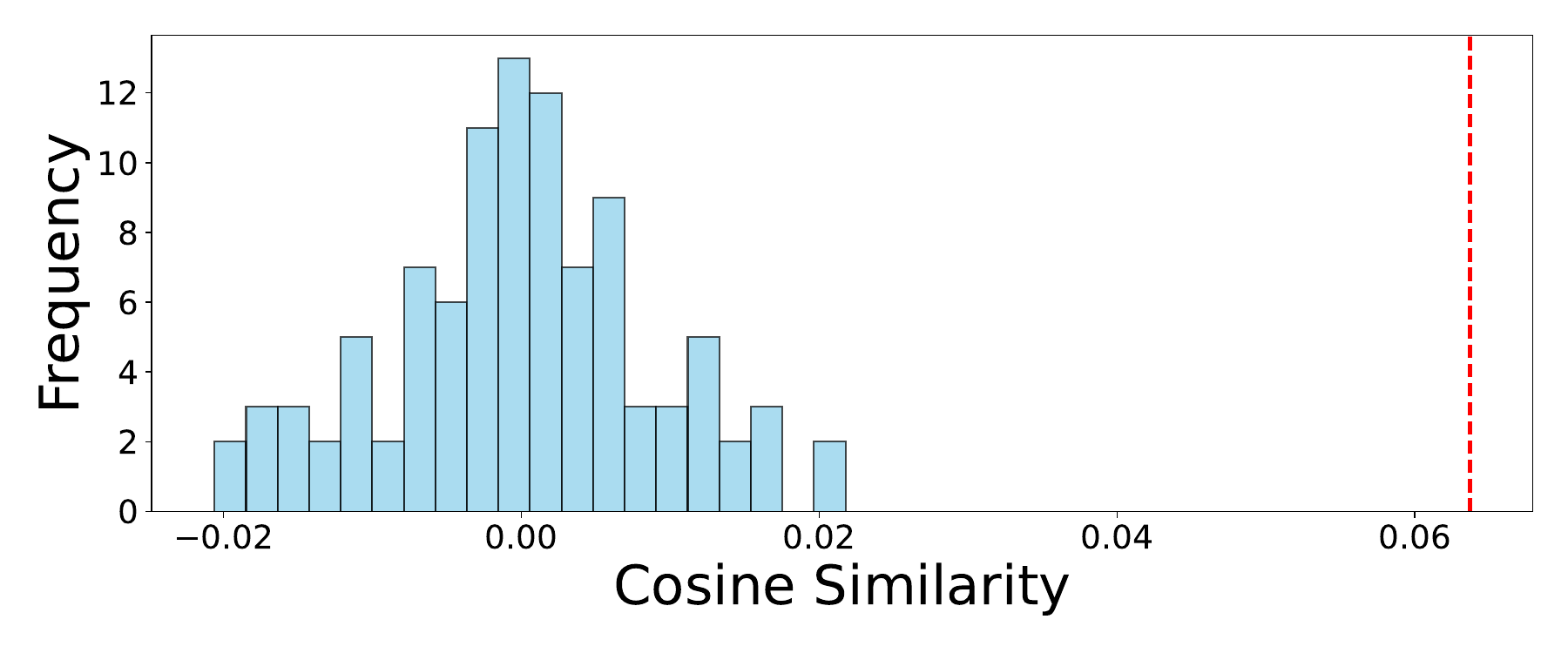}

    \caption{(Top) Distributions of random noise similarity
    versus originated noise similarity (red line) for the DiffWave vocoder [Left] and MatchTTS [Middle] and F5TTS models [Right]. The generated audio is significantly more similar to the original noise than random noises.}
    \label{fig:cosim_plot}
\end{figure*}

\subsubsection{F5-TTS and MatchaTTS}
For Text-to-Speech models, we use the official implementation of F5-TTS\footnote{https://github.com/SWivid/F5-TTS} and MatchaTTS \footnote{https://github.com/shivammehta25/Matcha-TTS}, a state-of-the-art Diffusion Transformer (DiT) based FM model \cite{Peebles_2023_ICCV}, with a similar pretrain task (text-guided
speech-infilling) as the E2 TTS model \cite{eskimez2024e2ttsembarrassinglyeasy}, allowing it to leverage reference audio for non-autoaggressive generation. On the other hand Matcha-TTS uses its own residual U-Net architecture, the training objective is more traditional, and it does not make use of reference audio, instead using speaker IDs for its multi-speaker TTS objective. For F5-TTS, we used their \textit{F5-TTS\_v1\_Base} checkpoint, with the Vocos vocoder \cite{siuzdak2024vocosclosinggaptimedomain}; and for Matcha-TTS, we used the VCTK checkpoints provided in their repository's ReadMe.

\subsubsection{Dataset}\label{model_data}
For the audio dataset, we choose the commonly used TTS datasets LibriTTS \cite{zen2019librittscorpusderivedlibrispeech} and LJSpeech \cite{ljspeech17}. LJSpeech is used for conducting experiments for the DiffWave vocoder.
In order to create the synthetic dataset to train our external detector, we used an additional emotion detection text dataset\footnote{https://huggingface.co/datasets/jakeazcona/short-text-labeled-emotion-classification}.

\subsubsection{Trained and Baseline Detector}
Our Audio NoisePrints detector is a 4-layer \textit{Conv2D} ResNet trained with a binary classification objective to detect whether a Mel-spectrogram was generated from our designated noise or not. During training we feed the model both Mel-spectrograms generated from randomly initialized noise and specified noise, and conduct binary classification using the BCE loss function.

We compare our model against the \textit{AudioSeal} \cite{roman2024proactivedetectionvoicecloning} baseline. It is a classic yet powerful post-hoc watermarking model, trained with 4.5K hours of data from the VoxPopuli dataset \cite{wang-etal-2021-voxpopuli}. It was used as a strong baseline in multiple post-hoc watermarking publications \cite{ liu2023detectingvoicecloningattacks, liu2024grootgeneratingrobustwatermark,kovacevic2025deepmark,özer2025comprehensiverealworldassessmentaudio}. We used their 16k pretrained checkpoint from the official repository \footnote{https://github.com/facebookresearch/audioseal/}.

\section{Results and Discussion}\label{result}

\subsection{Comparisons with AudioSeal}

In Figure \ref{fig:F5-TTS_audioseal_comparison}, \textit{AudioNoisePrints} is shown to be just as (if not more) robust against a trained baseline (AudioSeal) across F5-TTS and MatchaTTS, models with different architecture and training objective. While the top line accuracy does not reach 100\% (unlike Audioseal, but still reaching around 97-98\%), it performs significantly better under strong augmentations (except MP3, which does not seem to affect detection performance at all). On the other hand, \textit{AudioNoisePrints} significantly outperforms AudioSeal in cropping and speed augmentation, where AudioSeal completely failed with just a one percent change of audio speed. However, our method did not perform as well in the ``Echo" augmentation (see table \ref{table:compression_result}); this is not surprising, as AudioSeal is pretrained with this augmentation, while our method was not. Our detector produced strong results against conventional audio compression attacks ($>97\%$ precision and accuracy). Without any additional processing during generation or any compromise on generated quality, our model performs comparably to the commonly used \textit{AudioSeal} post-hoc watermarking model. 

The significance of the \textit{AudioNoisePrints} approach is its simplicity. While it might not be fair to compare a post-hoc watermarking scheme with an in-model (or in-generation) one, our new methodology has no computation overhead on the generation process or the need to retrain the TTS models (basically plug-and-play). This approach is also more robust than the post-hoc AudioSeal model in most augmentations, which not only specifically trained for the augmentations we tested, it also pretrained on a significantly larger dataset (4.5k hours of training data compared with us using just LibriTTS). This shows that if the watermark's objective is to protect (or detect) a Diffusion / FM Text-to-speech model, our method is much more computationally and data efficient.

\subsection{The choice of spatial distance function}
We also experimented with three other distance functions (L1, L2 and dot product). In Table \ref{tab:similarity_thresholds}, \textit{only} cosine similarity and dot-product are able to achieve any acceptable performance, though cosine similarity perform slightly better, with \textit{cutoff}$=1.0$ having the highest accuracy. Accordingly, all the following experiments will use $CosSim$ as their distance function.

\subsection{Spatial correlations in other TTS models}\label{spatial_correlation_exp}


As additional validation, we also analyze another diffusion vocoder. we evaluated \textit{DiffWave} \cite{kong2021diffwaveversatilediffusionmodel}, a diffusion vocoder. DiffWave uses a bidirectional dilated convolution architecture with diffusion modeling, a different architecture than Matcha-TTS or F5-TTS.
We used the official implementation for both MatchaTTS\footnote{https://github.com/shivammehta25/Matcha-TTS} and DiffWave\footnote{https://github.com/lmnt-com/diffwave/} in all of our experiments. For Matcha-TTS, we used the VCTK checkpoints; and for DiffWave we used the LJSpeech checkpoint.

In Figure \ref{fig:cosim_plot}, we can see the distance distribution of $d_{rand}$ (in blue) versus the original distance $d_{org}$ (in red) in a single audio sample (and with different models, namely, F5-TTS, Matcha-TTS, and the DiffWave vocoder). We can immediately see the large divide between $d^{rand}$ and $d_{org}$.  We further generate 500 audio data for MatchaTTS and DiffWave, and in every single audio, the p-value of the red line is 0.0, meaning $d_{org} >> d^{rand}$ is a general phenomenon for both models. The finding above shows the spatial-correlation is not unique to the TTS model we tested, suggesting similar watermarking and detection methodology could be adapted to other FM TTS models and vocoders. 


\section{Conclusions}
We present \textit{AudioNoisePrints}, a model-free watermarking scheme for fast audio watermarking. Comparing with the commonly used \textit{post-hoc} watermarking scheme such as \textit{AudioSeal}, our methodology is model-free, with zero computational overhead during generation, and without compromising the generation quality with the trade-off of robustness. Our experiments also show that \textit{AudioNoisePrints} is more robust under strong augmentations compared to the baseline \textit{AudioSeal}, which is trained with thousands of hours of audio data. We show that the spatial correlation property that enabled our watermarking methodology also exists in other FM and diffusion models. These findings suggest potential for broader adaptations and applications in different audio models.

\section{Use of Generative AI Disclosure}
No generative AI tools were used in the process of generating experiment results or writing this manuscript.
\section{Acknowledgments}
This work was done during first author's internship at the Digital Technologies Research Centre, National Research Council Canada. A special thanks to Robin Shing-Hei Yuen for providing computation resources when needed.

\bibliographystyle{IEEEtran}
\bibliography{mybib}

@inproceedings{
lipman2023flow,
title={Flow Matching for Generative Modeling},
author={Yaron Lipman and Ricky T. Q. Chen and Heli Ben-Hamu and Maximilian Nickel and Matthew Le},
booktitle={The Eleventh International Conference on Learning Representations },
year={2023},
url={https://openreview.net/forum?id=PqvMRDCJT9t}
}

@inproceedings{NEURIPS2020_4c5bcfec,
 author = {Ho, Jonathan and Jain, Ajay and Abbeel, Pieter},
 booktitle = {Advances in Neural Information Processing Systems},
 editor = {H. Larochelle and M. Ranzato and R. Hadsell and M.F. Balcan and H. Lin},
 pages = {6840--6851},
 publisher = {Curran Associates, Inc.},
 title = {Denoising Diffusion Probabilistic Models},
 url = {https://proceedings.neurips.cc/paper_files/paper/2020/file/4c5bcfec8584af0d967f1ab10179ca4b-Paper.pdf},
 volume = {33},
 year = {2020}
}

@misc{chen2025f5ttsfairytalerfakesfluent,
      title={F5-TTS: A Fairytaler that Fakes Fluent and Faithful Speech with Flow Matching}, 
      author={Yushen Chen and Zhikang Niu and Ziyang Ma and Keqi Deng and Chunhui Wang and Jian Zhao and Kai Yu and Xie Chen},
      year={2025},
      eprint={2410.06885},
      archivePrefix={arXiv},
      primaryClass={eess.AS},
      url={https://arxiv.org/abs/2410.06885}, 
}

@misc{mehta2024matchattsfastttsarchitecture,
      title={Matcha-TTS: A fast TTS architecture with conditional flow matching}, 
      author={Shivam Mehta and Ruibo Tu and Jonas Beskow and Éva Székely and Gustav Eje Henter},
      year={2024},
      eprint={2309.03199},
      archivePrefix={arXiv},
      primaryClass={eess.AS},
      url={https://arxiv.org/abs/2309.03199}, 
}

@misc{eskimez2024e2ttsembarrassinglyeasy,
      title={E2 TTS: Embarrassingly Easy Fully Non-Autoregressive Zero-Shot TTS}, 
      author={Sefik Emre Eskimez and Xiaofei Wang and Manthan Thakker and Canrun Li and Chung-Hsien Tsai and Zhen Xiao and Hemin Yang and Zirun Zhu and Min Tang and Xu Tan and Yanqing Liu and Sheng Zhao and Naoyuki Kanda},
      year={2024},
      eprint={2406.18009},
      archivePrefix={arXiv},
      primaryClass={eess.AS},
      url={https://arxiv.org/abs/2406.18009}, 
}

@misc{kong2021diffwaveversatilediffusionmodel,
      title={DiffWave: A Versatile Diffusion Model for Audio Synthesis}, 
      author={Zhifeng Kong and Wei Ping and Jiaji Huang and Kexin Zhao and Bryan Catanzaro},
      year={2021},
      eprint={2009.09761},
      archivePrefix={arXiv},
      primaryClass={eess.AS},
      url={https://arxiv.org/abs/2009.09761}, 
}

@misc{luo2025wavefmhighfidelityefficientvocoder,
      title={WaveFM: A High-Fidelity and Efficient Vocoder Based on Flow Matching}, 
      author={Tianze Luo and Xingchen Miao and Wenbo Duan},
      year={2025},
      eprint={2503.16689},
      archivePrefix={arXiv},
      primaryClass={cs.SD},
      url={https://arxiv.org/abs/2503.16689}, 
}

@misc{staniszewski2025againrelationnoiseimage,
      title={There and Back Again: On the relation between Noise and Image Inversions in Diffusion Models}, 
      author={Łukasz Staniszewski and Łukasz Kuciński and Kamil Deja},
      year={2025},
      eprint={2410.23530},
      archivePrefix={arXiv},
      primaryClass={cs.CV},
      url={https://arxiv.org/abs/2410.23530}, 
}

@misc{goren2025noiseprintsdistortionfreewatermarksauthorship,
      title={NoisePrints: Distortion-Free Watermarks for Authorship in Private Diffusion Models}, 
      author={Nir Goren and Oren Katzir and Abhinav Nakarmi and Eyal Ronen and Mahmood Sharif and Or Patashnik},
      year={2025},
      eprint={2510.13793},
      archivePrefix={arXiv},
      primaryClass={cs.CV},
      url={https://arxiv.org/abs/2510.13793}, 
}

@inproceedings{
choi2025visual,
title={Visual Fidelity vs. Robustness: Trade-Off Analysis of Image Adversarial Watermark Mitigated by {SSIM} Loss},
author={Jiwoo Choi and Jinwoo Kim and Sejong Yang and Seon Joo Kim},
booktitle={The 1st Workshop on GenAI Watermarking},
year={2025},
url={https://openreview.net/forum?id=MS7QPrBngC}
}

@misc{ho2020denoisingdiffusionprobabilisticmodels,
      title={Denoising Diffusion Probabilistic Models}, 
      author={Jonathan Ho and Ajay Jain and Pieter Abbeel},
      year={2020},
      eprint={2006.11239},
      archivePrefix={arXiv},
      primaryClass={cs.LG},
      url={https://arxiv.org/abs/2006.11239}, 
}

@misc{lai2025principlesdiffusionmodels,
      title={The Principles of Diffusion Models}, 
      author={Chieh-Hsin Lai and Yang Song and Dongjun Kim and Yuki Mitsufuji and Stefano Ermon},
      year={2025},
      eprint={2510.21890},
      archivePrefix={arXiv},
      primaryClass={cs.LG},
      url={https://arxiv.org/abs/2510.21890}, 
}

@misc{roman2024proactivedetectionvoicecloning,
      title={Proactive Detection of Voice Cloning with Localized Watermarking}, 
      author={Robin San Roman and Pierre Fernandez and Alexandre Défossez and Teddy Furon and Tuan Tran and Hady Elsahar},
      year={2024},
      eprint={2401.17264},
      archivePrefix={arXiv},
      primaryClass={cs.SD},
      url={https://arxiv.org/abs/2401.17264}, 
}

@misc{chen2024wavmarkwatermarkingaudiogeneration,
      title={WavMark: Watermarking for Audio Generation}, 
      author={Guangyu Chen and Yu Wu and Shujie Liu and Tao Liu and Xiaoyong Du and Furu Wei},
      year={2024},
      eprint={2308.12770},
      archivePrefix={arXiv},
      primaryClass={cs.SD},
      url={https://arxiv.org/abs/2308.12770}, 
}

@misc{roman2024latentwatermarkingaudiogenerative,
      title={Latent Watermarking of Audio Generative Models}, 
      author={Robin San Roman and Pierre Fernandez and Antoine Deleforge and Yossi Adi and Romain Serizel},
      year={2024},
      eprint={2409.02915},
      archivePrefix={arXiv},
      primaryClass={cs.SD},
      url={https://arxiv.org/abs/2409.02915}, 
}

@misc{liu2023detectingvoicecloningattacks,
      title={Detecting Voice Cloning Attacks via Timbre Watermarking}, 
      author={Chang Liu and Jie Zhang and Tianwei Zhang and Xi Yang and Weiming Zhang and Nenghai Yu},
      year={2023},
      eprint={2312.03410},
      archivePrefix={arXiv},
      primaryClass={cs.SD},
      url={https://arxiv.org/abs/2312.03410}, 
}

@misc{yao2025mineoverwritingattacksneural,
      title={Yours or Mine? Overwriting Attacks against Neural Audio Watermarking}, 
      author={Lingfeng Yao and Chenpei Huang and Shengyao Wang and Junpei Xue and Hanqing Guo and Jiang Liu and Phone Lin and Tomoaki Ohtsuki and Miao Pan},
      year={2025},
      eprint={2509.05835},
      archivePrefix={arXiv},
      primaryClass={cs.CR},
      url={https://arxiv.org/abs/2509.05835}, 
}

@misc{liu2024grootgeneratingrobustwatermark,
      title={GROOT: Generating Robust Watermark for Diffusion-Model-Based Audio Synthesis}, 
      author={Weizhi Liu and Yue Li and Dongdong Lin and Hui Tian and Haizhou Li},
      year={2024},
      eprint={2407.10471},
      archivePrefix={arXiv},
      primaryClass={cs.CR},
      url={https://arxiv.org/abs/2407.10471}, 
}

@misc{zhao2025traceablettswatermarkfreetts,
      title={Traceable TTS: Toward Watermark-Free TTS with Strong Traceability}, 
      author={Yuxiang Zhao and Yunchong Xiao and Yushen Chen and Zhikang Niu and Shuai Wang and Kai Yu and Xie Chen},
      year={2025},
      eprint={2507.03887},
      archivePrefix={arXiv},
      primaryClass={eess.AS},
      url={https://arxiv.org/abs/2507.03887}, 
}

@misc{zen2019librittscorpusderivedlibrispeech,
      title={LibriTTS: A Corpus Derived from LibriSpeech for Text-to-Speech}, 
      author={Heiga Zen and Viet Dang and Rob Clark and Yu Zhang and Ron J. Weiss and Ye Jia and Zhifeng Chen and Yonghui Wu},
      year={2019},
      eprint={1904.02882},
      archivePrefix={arXiv},
      primaryClass={cs.SD},
      url={https://arxiv.org/abs/1904.02882}, 
}

@misc{ljspeech17,
  author       = {Keith Ito and Linda Johnson},
  title        = {The LJ Speech Dataset},
  howpublished = {\url{https://keithito.com/LJ-Speech-Dataset/}},
  year         = 2017
}

@InProceedings{Peebles_2023_ICCV,
    author    = {Peebles, William and Xie, Saining},
    title     = {Scalable Diffusion Models with Transformers},
    booktitle = {Proceedings of the IEEE/CVF International Conference on Computer Vision (ICCV)},
    month     = {October},
    year      = {2023},
    pages     = {4195-4205}
}

@inproceedings{wang-etal-2021-voxpopuli,
    title = "{V}ox{P}opuli: A Large-Scale Multilingual Speech Corpus for Representation Learning, Semi-Supervised Learning and Interpretation",
    author = "Wang, Changhan  and
      Riviere, Morgane  and
      Lee, Ann  and
      Wu, Anne  and
      Talnikar, Chaitanya  and
      Haziza, Daniel  and
      Williamson, Mary  and
      Pino, Juan  and
      Dupoux, Emmanuel",
    booktitle = "Proceedings of the 59th Annual Meeting of the Association for Computational Linguistics and the 11th International Joint Conference on Natural Language Processing (Volume 1: Long Papers)",
    month = aug,
    year = "2021",
    address = "Online",
    publisher = "Association for Computational Linguistics",
    url = "https://aclanthology.org/2021.acl-long.80",
    pages = "993--1003",
}

@inproceedings{
kovacevic2025deepmark,
title={DeepMark Benchmark: Redefining Audio Watermarking Robustness},
author={Slavko Kova{\v{c}}evi{\'c} and Murilo Z. Silvestre and Kosta Pavlovi{\'c} and Petar Nedi{\'c} and Igor Djurovi{\'c}},
booktitle={The 1st Workshop on GenAI Watermarking},
year={2025},
url={https://openreview.net/forum?id=56ZC5dqvJO}
}

@misc{özer2025comprehensiverealworldassessmentaudio,
      title={A Comprehensive Real-World Assessment of Audio Watermarking Algorithms: Will They Survive Neural Codecs?}, 
      author={Yigitcan Özer and Woosung Choi and Joan Serrà and Mayank Kumar Singh and Wei-Hsiang Liao and Yuki Mitsufuji},
      year={2025},
      eprint={2505.19663},
      archivePrefix={arXiv},
      primaryClass={cs.SD},
      url={https://arxiv.org/abs/2505.19663}, 
}

@misc{siuzdak2024vocosclosinggaptimedomain,
      title={Vocos: Closing the gap between time-domain and Fourier-based neural vocoders for high-quality audio synthesis}, 
      author={Hubert Siuzdak},
      year={2024},
      eprint={2306.00814},
      archivePrefix={arXiv},
      primaryClass={cs.SD},
      url={https://arxiv.org/abs/2306.00814}, 
}

\end{document}